\documentclass[letterpaper]{article}
\usepackage{amsmath}
\usepackage{dsfont}
\usepackage{aaai}
\usepackage{times}
\usepackage{helvet}
\usepackage{courier}
\usepackage{color}
\usepackage{graphicx}
\usepackage{multicol}
\usepackage{balance}
\usepackage{amsmath}
\usepackage{amsfonts}

\newcommand{\laCTR}{{$\mathcal LA$}-CTR}

\newcommand{\commentout}[1]{%
}

\newcommand{\figref}[1]{Figure~\ref{#1}}

\newcommand{\remove}[1]{}
\begin{document}
 \bibliographystyle{aaai}
%
\title{LA-CTR:
A Limited Attention Collaborative Topic Regression for Social Media}
\author{
Jeon-Hyung Kang \\
USC Information Sciences Institute\\
4676 Admiralty Way \\
Marina del Rey, CA 90292 \\
\texttt{jeonhyuk@usc.edu} \\
\And
Kristina Lerman \\
USC Information Sciences Institute\\
4676 Admiralty Way \\
Marina del Rey, CA 90292 \\
\texttt{lerman@isi.edu}
}
\maketitle

\begin{abstract}
Probabilistic models can learn users' preferences from the history of their item adoptions on a social media site, and in turn, recommend new items to users based on learned preferences. However, current models ignore psychological factors that play an important role in shaping online social behavior. One such factor is attention, the mechanism that integrates perceptual and cognitive features to select the items the user will consciously process and may eventually adopt. Recent research has shown that people have finite attention, which constrains their online interactions, and that they divide their limited attention non-uniformly over other people. We propose a collaborative topic regression model that incorporates limited, non-uniformly divided attention. We show that the proposed model is able to learn more accurate user preferences than state-of-art models, which do not take human cognitive factors into account. Specifically we analyze voting on news items on the social news aggregator and show that our model is better able to predict held out votes than alternate models. Our study demonstrates that psycho-socially motivated models are better able to describe and predict observed behavior than models which only consider latent social structure and content.
\end{abstract}

\section{Introduction}

The rise of social media has drastically exacerbated the problem of information overload. Social media users create far more information than other users can process. Social media sites address this problem by allowing users to follow updates from specified friends; however, as the number of friends a user follows grows over time, information overload returns. One solution to this problem is to recommend items to a user based on his or her interests or preferences, which can be learned by observing user's activity. Researchers have developed  statistical models to learn users' preferences from the items they, and their friends, have adopted in social media~\cite{Lauw12,purushotham}, and filter the streams of new information accordingly.

Existing statistical models largely ignore the psychological and social factors that shape user behavior. However, recent research has demonstrated that such factors, in particular attention, play an important role in social media users' behavior and interactions~\cite{Counts11,Hodas12socialcom,Goncalves11,Weng:2012dd}. Attention is the mechanism that integrates perceptual and cognitive factors to select the small fraction of input to be processed in real time~\cite{Kahneman73,Rensink:1997vj}. Attentive acts, such as reading a tweet, browsing a web page, or responding to email, require mental effort, and since the brain's capacity for mental effort is limited, so is attention. Limited attention was shown to constrain user's social interactions~\cite{Goncalves11} and the spread of information~\cite{Hodas12socialcom} in social media. Moreover, while users divide their limited attention over their friends~\cite{Hodas12socialcom}, some friends receive a larger share of their attention than others~\cite{Gilbert09}.

In this paper, we close the gap between behavioral and statistical recommendation models by presenting {\laCTR}, a model that extends collaborative recommendation topic model CTR introduced by Wang \& Blei~\cite{WangB11} by integrating users' limited attention into it. Like CTR, {\laCTR} uses item adoptions of all users and the content of items as a basis for its recommendations. However, {\laCTR} modifies the existing model in two important ways. First, the model captures a user's limited attention, which he or she must divide among all others she follows. Second, the model includes a notion of influence, that is the fact that the user may preferentially pay more attention to some of the users, e.g., close friends~\cite{Gilbert09}. Moreover, the model allows the influence to vary according to each user's topical interests. We evaluate the model on real world data from the social news site Digg. We demonstrate that the proposed psycho-socially motivated model performs better on the news story recommendation task than alternative models.

In the rest of the paper we first review existing research, which  includes two state-of-art probabilistic collaborative models: CTR and CTR-smf. In Section ``{\laCTR}'', we introduce a limited attention collaborative regression model that takes into consideration the limited, non-uniformly divided attention of social media users. In Section ``{Evaluation}'', we study {\laCTR} model with votes on news items on the social news aggregator Digg  and show that our proposed model is better able to predict held out votes than alternative models that do not take limited attention into account. Our study demonstrates that behaviorally motivated models are better able to describe observed user behavior in social media, and  lead to better recommendation tools.

\section{Related Work}
\label{sec:relatedwork}

The growing abundance of social media data has allowed researchers to start asking questions about the nature of influence, how social ties interact with the contents of items, and how these affect transmission of items in social networks~\cite{Bakshy11,Romero11www}. Recently, researchers demonstrated the importance of attention in online interactions~\cite{Goncalves11,Hodas12socialcom,Weng:2012dd}. Specifically relevant to this paper are studies that show that limited attention constrains social interactions~\cite{Goncalves11} and information diffusion~\cite{Hodas12socialcom} in social media. The model proposed  in this paper incorporates these ideas into a hidden topic model for collaborative recommendation.

In text mining applications, hidden topic models, such as Latent Dirichlet Allocation (LDA)~\cite{blei2003latent}, examine documents to identify hidden topics, each represented as distributions over groups of related words. Traditional topic models have been extended to a networked setting to model hyperlinks between documents~\cite{citeulike:2914131}, topical relations between them~\cite{Chang_relationaltopic,DaumŽ09markovrandom}, and the varying vocabularies and styles of different authors~\cite{RosenZvtheauthortopic}.
More generally, researchers have applied topic models to social media analysis, e.g., by treating users as documents and content they create as words~\cite{ramage2009labeled}. Then, by analyzing messages posted by many users, one can separate users into topical groups.

Collaborative filtering methods examine item recommendations of many people to discover their preferences and recommend new items that were liked by similar people. User-based~\cite{herlocker1999algorithmic} and item-based~\cite{Sarwar01itembasedcollaborative,karypis2000evaluation} recommendation approaches have been developed for predicting how people will rate new items.  Matrix factorization-based latent-factor models~\cite{salakhutdinov2008probabilistic,koren2009matrix} have shown promise in creating better recommendations by incorporating user interests into the model. However all these approaches ignore the content of items in the recommendation task.
Recently, probabilistic topic modeling techniques were merged with collaborative filtering to improve the explanatory power of recommendation tools~\cite{AgarwalC10,WangB11}.
Content analysis based on probabilistic topic modeling has been proposed to incorporate into collaborative filtering ~\cite{AgarwalC10} to account for user rating and item popularity biases in the data.  Authors show better prediction performance by regularization of  both user and item factors through user features and the content of item.
Collaborative Topic Regression (CTR) model~\cite{WangB11} incorporates collaborative filtering model based on probabilistic topic modeling.
Both models do a good job in using item content for recommendation; however, neither takes social structure of users into account.

Researchers have extended LDA to a social recommendation setting~\cite{Lauw12},  in which a user is likely to adopt an item based on the latent interests of other users. Social correlations between users are learned from their item adoption choices, rather than specified explicitly through the follower graph. Social recommendation system has been proposed by matrix factorization techniques for both user's social network and their item rating histories~\cite{ma2008sorec,yang2011like} without using the contents of items. In CTR-smf~\cite{purushotham}, authors integrate CTR with social matrix factorization models to take social correlation between users into account. However, these works utilize \emph{homophily} effect in social media to smooth the similarity of users' interests to their friends, instead of directly learning how users allocate their  attention over their friends. So depending on the degree of similarity among connected users, social matrix factorization may not consistently perform well.
The model proposed in this paper directly learns how much attention users allocate to other users, and leverages these learned influences to make better recommendations.

Attention in online interactions has been modeled in a previous work~\cite{Kang13sbp} to directly learn users' limited attention over their friends and topics. However, that model was applied to dyadic data~\cite{Agarwal2009,koren2009matrix,Yu2009} without providing any explanation about the learned topics.
Furthermore, authors assumed a simple user item adoption mechanism, where item adoption is decided based on sampled interest and topic pairs, rather than the inner product of user interest and item topic latent vectors, as matrix factorization does. The model proposed in this paper makes two important advances. First, it incorporates the content of items, therefore, being able to provide explanatory power to its recommendations, as well as recommend previously unseen items. Furthermore, the proposed model is able to learn the degree of influence of other users in modeling limited attention.

\section{\laCTR: Limited Attention Collaborative Topic Regression Model}
\label{sec:lactr}
\begin{figure}[b]
\begin{center}
\includegraphics[width=0.7\linewidth]{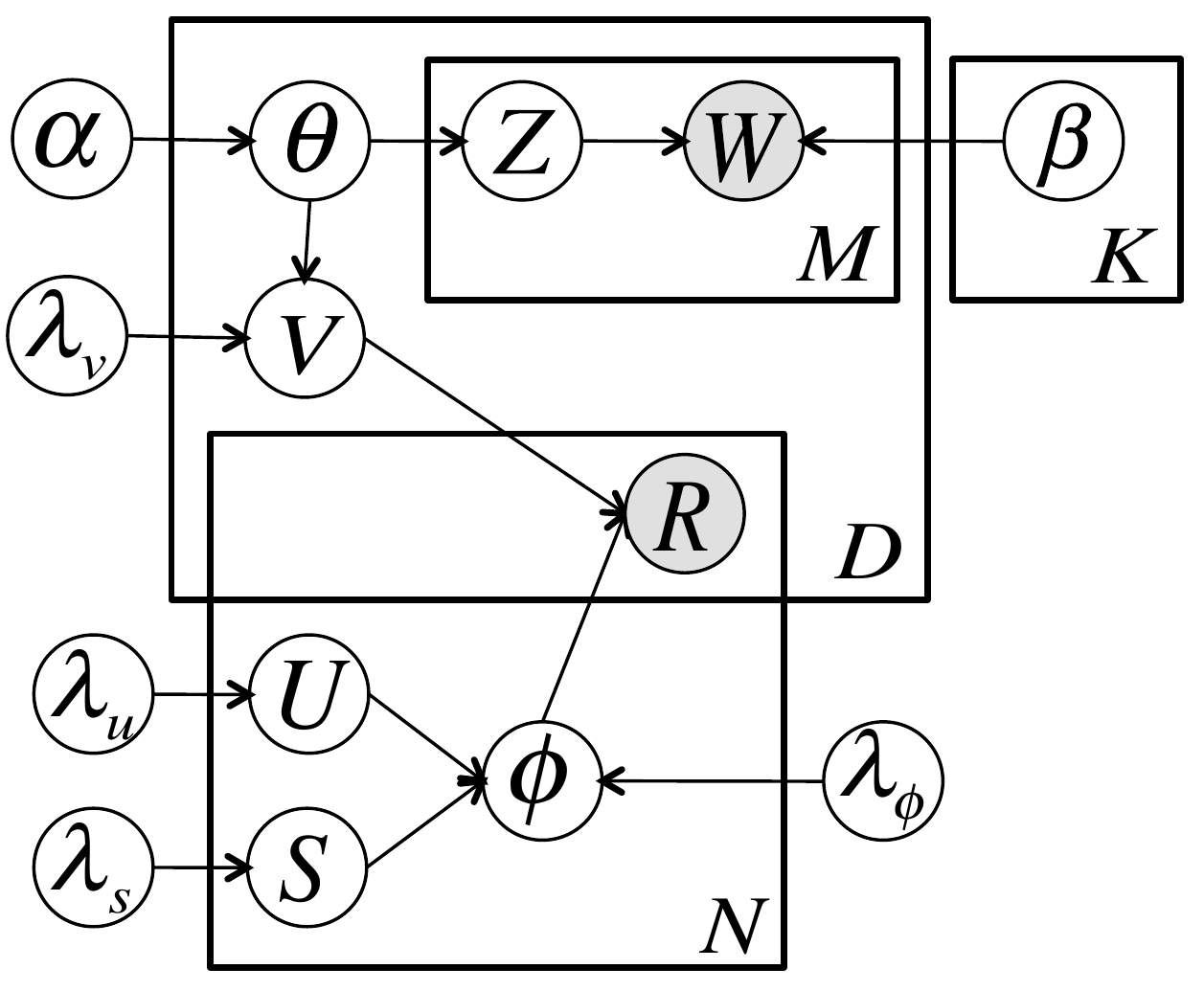}
\end{center}
\caption{The {\laCTR} model.}
\label{fig:modelDiagram}
\end{figure}
We introduce a limited attention collaborative topic regression model (\laCTR) that learns user preferences from the observations of items they adopt on a social media site. Our model captures the behavior of social media users by introducing  three novel elements that model users' limited attention ($\phi$), their interests ($u$), and how much attention they pay to other users ($s$), which we also refer to as influence. Two additional elements in the model are the item's topic for explaining recommendations ($v$) and for explaining the item's contents ($\theta$), which were introduced by \cite{WangB11}. Finally ratings ($r$) are observed variables. Rating is equal to one when user adopts an item and zero otherwise.

Figure~\ref{fig:modelDiagram} presents  {\laCTR} in graphical form. Each variable is drawn from a normal distribution defined as follows:
\begin{equation}
\begin{aligned}
& u_{i} \sim \mathcal{N}(0,  \lambda_{u}^{-1}I_{K})\\
& s_{i} \sim \mathcal{N}(0,  \lambda_{s}^{-1}I_{N})\\
& \phi_{il} \sim \mathcal{N}\left(g_\phi({s_{il} {u_i}}),c_{il}^{\phi}\lambda_{\phi}^{-1}I_{K}\right) \\
& r_{ijl} \sim {\mathcal{N} \left(  g_r(\phi^T_{il} v_j), c_{ijl}^{r}  \right) } \\
 \end{aligned}
\end{equation}
where $N$ is the number of users, ${D}$ is the number of items, and $K$ is the number of topics. Symbol $T$ refers to the transpose operation.
We define $g_\phi$ and $g_r$ as linear functions for simplicity.
The precision parameters $c_{il}^{\phi}$ and $c_{ijl}^{r}$ serve as confidence parameters for influence $s_{il}$ and rating $r_{ijl}$.
The  confidence parameter  $c_{il}^{\phi}$ represents how much {attention user $i$ pays to user $l$}. In  social  media, users  adopt new items primarily based on their friends' recommendations, i.e., by seeing the items their friends adopt. However, users may also  be exposed to items from outside  of their social network, depending on the site's interface.  We use $c_{il}^{\phi}$ to model the structure of the social network:  we set it a high value $a^\phi$ when user $l$ is a friend of user $i$ and a low value $b^\phi$ when he is not ($a^\phi>b^\phi>0$).
Similarly, when user $i$ does not adopt item $j$ (i.e., her rating of item $j$ is zero), it can be interpreted into two ways: either user $i$ was aware of item $j$ but did not like it, or user $i$ would have liked item $j$ but never saw it~\cite{WangB11}. We set $c_{ijl}^{r}$ as a confidence parameter for user $i$'s rating on item $j$ via user $l$ and set it to a high value $a^r$ when $r_{ijl}=1$ and a low value $b^r$ when $r_{ijl}=0$ ($a^r>b^r>0$). In this paper, we use the same confidence parameter values, $a^r=a^\phi=1.0$ and $b^r=b^\phi=0.01$, for all $c_{il}^{\phi}$ and $c_{ijl}^{r}$.

Following CTR, we use the hidden topic model LDA~\cite{blei2003latent} to capture item's topic distribution $\theta$, which is represented as a $K$ dimensional vector. In LDA, the topic distribution ($\theta_{1:D}$) of each document is viewed as a  mixture of multiple topics, with each topic ($\beta_{1:K}$) a distribution over words from a vocabulary of size $M$.  Like CTR, we assume a generative process in which documents are created by drawing words from their topic distributions. Latent variable $\epsilon_j \sim \mathcal{N}(0,\lambda_{v}^{-1}I_{K})$ captures the differences between topics that explain the contents of documents and those that explain recommendations. Depending on the choice of $\lambda_v$, item $j$'s topic  distribution $\theta_j$ is perturbed to create latent vector $v_j$, which could be similar to $\theta_j$ or diverge from it:
\begin{equation}
\begin{aligned}
& v_{j} \sim \mathcal{N}(\theta_j,  \lambda_{v}^{-1}I_{K})\\
 \end{aligned}
\end{equation}

The generative process for item adoption through a social network can be formalized as follows:
\begin{tabbing}
For \=each user $i$ \\
\>  Generate \= $u_{i} \sim \mathcal{N}(0,\lambda_{u}^{-1}I_{K}$)\\
\>  Generate \= $s_{i} \sim \mathcal{N}(0,\lambda_{s}^{-1}I_{N}$)\\
\>  For \=each user $l$ \\
\>\;\;\;\;\;\;Generate \= $\phi_{il} \sim \mathcal{N}(g_\phi({s_{il} {u_i}}),c_{il}^{\phi}\lambda_{\phi}^{-1}I_{K}$)\\
For \=each item j \\
\>  Generate \= $\theta_{j} \sim Dirichlet({\alpha}$) \\
\>  Generate \= $\epsilon_{j} \sim \mathcal{N}(0,\lambda_{v}^{-1}I_{K}$) and set \= $v_{j} = \epsilon_{j} + \theta_{j}$\\
\>\;\;\;\;\;\;For each word $w_{jm}$ \\
\>\>Generate topic assignment $z_{jm} \sim Mult ( \theta$)\\
\>\> Generate word $w_{jm} \sim Mult ( \beta_{z_{jm} }$)\\
For \=each  user $i$\\
\> For each attention friend $l$\\
\>\;\;\;\;\;\;For each adopted item $j$\\
\>\> Choose the rating $r_{ijl} \sim \mathcal{N}( \phi^T_{il} v_j , {c_{ijl}^{r}}^{-1}$)\\
\end{tabbing}
Here $\lambda_u = \sigma_r^2/\sigma_u^2$, $\lambda_v$ = $\sigma_r^2/\sigma_v^2$, $\lambda_s = \sigma_r^2/\sigma_s^2$, and $\lambda_\phi$ = $\sigma_r^2/\sigma_\phi^2$. Note that latent vectors $u$, $v$ and $\phi$ are in a shared $K$-dimensional space.

\subsection{Learning Parameters}
To learn model parameters, we follow the approaches of CTR~\cite{WangB11} and CTR-smf~\cite{purushotham} and develop an EM-style algorithm to calculate the maximum a posteriori (MAP) estimates. MAP estimation is equivalent to maximizing the complete log likelihood ($\ell$) of $U$, $S$, $\phi$, $V$, ${R}$, and $\theta$ given $\lambda_u$, $\lambda_s$, $\lambda_\phi$, $\lambda_v$ and $\beta$. We set Dirichlet prior $\alpha$ to 1.

\begin{equation}
\begin{aligned}
\ell =
& -\frac{\lambda_u}{2} \sum_i^N u_i^T u_i -\frac{\lambda_v}{2} \sum_j^D {(v_j - \theta_j)}^T {(v_j - \theta_j)} \\
& +\sum_j^D\sum_t^{W(j)} \log\left(\sum_k^K \theta_{jk} \beta_{k,w_{jt}}\right) -\frac{\lambda_s}{2} \sum_i^N {s_{i}}^T {s_{i}} \\
& -\sum_i^N\sum_j^D\sum_l^{N}  \frac{c_{ijl}^r}{2} {(r_{ijl} - { \phi_{il}^T v_j} )}^2 \\
&-\frac{\lambda_\phi}{2}  \sum_i^N\sum_l^{N} c_{il}^\phi {(\phi_{il} - {s_{il} {u_i}} )}^T (\phi_{il} - {s_{il} {u_i}} )
 \end{aligned}
\end{equation}
We use gradient ascent to estimate MAP and iteratively optimize the variables \{$u_i, v_j, \phi_{il}, s_i$\} and the topic proportions $\theta_j$. Given a current estimate, we take the gradient of $\ell$ with respect to $u_i$, $v_j$, $s_i$, and $\phi_{il}$ and set it to zero. Derived update equations are following:
\begin{equation*}
\begin{aligned}
&u_i \leftarrow {\left( \lambda_u I_k + \lambda_{\phi} S_i^T C_i^\phi {S_i} \right)}^{-1} \lambda_\phi \phi_i C_i^\phi S_i  \\
&v_j \leftarrow {\left( \lambda_v I_k +  {\phi}  C_j^r {{\phi} }^T \right)}^{-1}
\left( {\phi}  C_j^r R_j + \lambda_v  \theta_j \right)\\
&S_{i} \leftarrow {\left( \lambda_s I_N + \lambda_\phi {u_i}^T C_{i}^\phi {u_i} \right)}^{-1}  \lambda_\phi  C_{i}^\phi \phi_{i}^T{u_i} \\
&\phi_{il} \leftarrow {\left( \lambda_\phi C_{il}^\phi I_K + V C_{il}^r V^T \right)}^{-1}
\left( V C_{il}^r R_{il} + \lambda_\phi {u_i} C_{il}^\phi  S_{il} \right)
 \end{aligned}
\end{equation*}
where $C_i^\phi$ and $C_{il}^r$ are diagonal matrices with confidence parameters $c_{ij}^\phi$ and $c_{ijl}^r$. We define $S$ as $N \times N$ matrix, $\phi$ as $K \times N^2$ matrix and $R_j$ as vector with $r_{ijl}$ values for all pairs of users $i$ and $l$ for the given item $j$. Given updated variables \{$u_i, v_j, \phi_{il}, s_i$\} the topic proportions $\theta_j$ is updated by applying Jensen's inequality \cite{WangB11}:
\begin{equation}
\begin{aligned}
\ell (\theta_j) \ge
& \sum_m^{W(j)}\sum_k^K \psi_{jmk} \left(\log \theta_{jk} \beta_{k,{w_{jm}}}- \log \psi_{jmk}\right) \\
&  -\frac{\lambda_v}{2} \sum_j^D {(v_j - \theta_j)}^T {(v_j - \theta_j)} \\
=& \ell(\theta_j,\psi_j)\\
 \end{aligned}
\end{equation}
where $\psi_{jmk} = q(z_{jm}=k)$ and $\ell(\theta_j,\psi_j)$ gives the tight lower bound of $\ell(\theta_j)$. The optimal $\psi_{jmk}$ satisfies $\psi_{jmk} \propto \theta_{jk} \beta_{k,{w_{jm}}}$. Given updated variables \{$u_i, v_j, \phi_{il}, s_i, \theta_j$\}, we can optimize $\beta$,
\begin{equation}
\begin{aligned}
\beta_{k\emph{w}} \propto  \sum_j\sum_m  \psi_{jmk} \delta(w_{jm} = \emph{w})
 \end{aligned}
\end{equation}
where $\delta$ is one if and only if $\emph{w}$ is assigned to $w_{jm}$.

\subsection{Prediction}
After all the parameters are learned, {\laCTR} model can be used for both in-matrix and out-of-matrix prediction with either user's attention ($\phi$) or interest ($u$). As \cite{WangB11} mentioned, in-matrix prediction refers to the prediction task for user's rating on item that has been rated at least once by other users, while out-of-matrix prediction refers to predicting user's rating on a new item that has no rating history.

For in-matrix prediction with attention ($\phi$), the prediction that user $i$'s ratings for item $j$ via user $l$ is obtained by point estimation with optimal variables ($\phi^{*}$, $u^{*}$, $v^{*}$, $\theta^{*}$):
\begin{equation}
\begin{aligned}
\mathbb{E}[r_{ijl}|\mathcal{D}] \approx&
     \mathbb{E}{[\phi_{il}|\mathcal{D}]}^T  (\mathbb{E}[\theta_{j}|\mathcal{D}] +\mathbb{E}[\epsilon_{j}|\mathcal{D}] )\\
r_{ij}^{*} \approx& {\phi_{il}^{*}}^T {v_{j}^{*}}
 \end{aligned}
\end{equation}
where the rating of user $i$ is decided by user's attention $\phi^{*}_{il}$ and item topic profile $v^{*}_j$.

For out-of-matrix prediction with attention ($\phi$), when the item is new and no ratings are observed, the point estimation equation is:
\begin{equation}
\begin{aligned}
\mathbb{E}[r_{ijl}|\mathcal{D}] \approx&
     \mathbb{E}{[\phi_{il}|\mathcal{D}]}^T  \mathbb{E}[\theta_{j}|\mathcal{D}] \\
r_{ij}^{*} \approx& {\phi_{il}^{*}}^T {\theta_{j}^{*}}
\end{aligned}
\end{equation}
where item topic profile is $\theta^{*}_j$ due to $\epsilon^{*}_j=0$.

For in-matrix prediction with user interest ($u$), the point estimation prediction that user $i$'s rating for item $j$ is:
\begin{equation}
\begin{aligned}
\mathbb{E}[r_{ij}|\mathcal{D}] \approx&
     \mathbb{E}{[u_{i}|\mathcal{D}]}^T  (\mathbb{E}[\theta_{j}|\mathcal{D}] +\mathbb{E}[\epsilon_{j}|\mathcal{D}] )\\
r_{ij}^{*} \approx& {u_{i}^{*}}^T {v_{j}^{*}}
 \end{aligned}
\end{equation}
where the prediction of user's rating is decided by user interest profile $u_i^*$ and item topic profile $v_j^*$.

For out-of-matrix prediction with user interest ($u$):
\begin{equation}
\begin{aligned}
\mathbb{E}[r_{ij}|\mathcal{D}] \approx&\mathbb{E}{[u_{i}|\mathcal{D}]}^T  \mathbb{E}[\theta_{j}|\mathcal{D}] \\
r_{ij}^{*} \approx& {u_{i}^{*}}^T {\theta_{j}^{*}}
 \end{aligned}
\end{equation}
where the rating of a user is decided by the user's interest profile $u_i^*$ and item's topic profile $\theta^{*}_j$.
\section{Evaluation}
\label{sec:evaluation}

In this section we demonstrate the utility of the {\laCTR} model both with interest ($u$) and attention ($\phi$) hidden variables. We apply the {\laCTR} model to data from the social news aggregator Digg. We compare our results to state-of-the-art alternative models: CTR and CTR-smf.

\subsection{Data Set}
Digg allows users to submit links to news stories and other users to recommend stories by voting for or ``digging'' stories they find interesting. Digg also allows users to follow the activity of other users to see the stories they submitted or voted for recently. When a user votes for a story, all her followers can see the vote, so the information is broadcast and shared with all social network neighbors. At the time datasets were collected, users were submitting tens of thousands of stories, from which Digg selected a handful (about 100) to promote to its front page based on how popular in the community.

We perform evaluations on two datasets: the 2009 dataset~\cite{Lerman10icwsm} and the 2010 dataset~\cite{sharara:icwsm11}. The 2009 dataset contains information about the voting history of 70K active users on 3.5K stories promoted to Digg front page in June 2009, and contains 2.1 million votes. The follower graph of contains 1.7 million social links. At the time this dataset was collected, Digg was assigning stories to one of 8 topics (Entertainment, Lifestyle, Science, Technology, World \& Business, Sports, Offbeat, and Gaming) and one of 50 subtopics (World News, Tech Industry News, General Sciences, Odd Stuff, Movies, Business, World News, Politics, etc.). The 2010 data set contains information about voting histories of 12K users over a 6 months period (Jul - Dec 2010). It includes 48K stories with 1.9 million votes and contains 1.3 million social links in the follower graph. At the time data was collected, Digg was assigning stories to 10 topics, replacing the 2009 ``World \& Business'' category with ``World News,'' ``Business,'' and ``Politics'' and dropping lower-level topic categories.

We examine only the votes that the story accrues before promotion to the front page.  During that time, the story propagates mainly via friends' votes (or recommendations), although some users could discover the story on the Upcoming stories page, which received tens of thousands of submissions daily.
 After promotion, users are likely to be exposed to the story through the front page, and vote for it independently of friends' recommendations. In the 2009 data set, 28K users voted for 3,553 stories and in the 2010 data set, 4K users voted for 36,883 stories before promotion.
We use tf-idf to choose the top 3K distinct words from the titles and user-supplied summaries of stories in the 2009 data set and 4K distinct words in the 2010 data set as the vocabularies. We focused the data further by selecting only those users who voted at least 10 times and items containing more than 10 words, resulting in 1.4K users (who voted for 3K stories) in the 2009 data set and 1.8K users (who voted on 18K stories) in the 2010 data set.

\subsection{Evaluation on Vote Prediction}
We consider a vote by user $i$ on story $j$ as an item adoption event, or a rating of one.
A rating of zero means either that the user did not like the item or was not aware of it; therefore, we only consider rated items for evaluation. For a fair and easy comparison with CTR and CTR-smf, we use $recall@\emph{X}$.
We present each user with $\emph{X}$  items sorted by their predicted rating score, and evaluate based on the fraction of items that the user actually voted for. We define $recall@\emph{X}$ for a user as:
\begin{equation}
\begin{aligned}
recall@\emph{X} =
& \frac{\text{num of items in top \emph{X} user votes for}}{\text{total num of items user votes for}} \\
 \end{aligned}
\end{equation}
and we average recall values of all users to summarize performance of the algorithm.  Note a better model should provide higher $recall@\emph{X}$ at each $\emph{X}$.

\subsection{Impact of Parameters}

\begin{figure}[tbh]
\begin{center}
\begin{tabular}{c}
\includegraphics[width=0.8\linewidth]{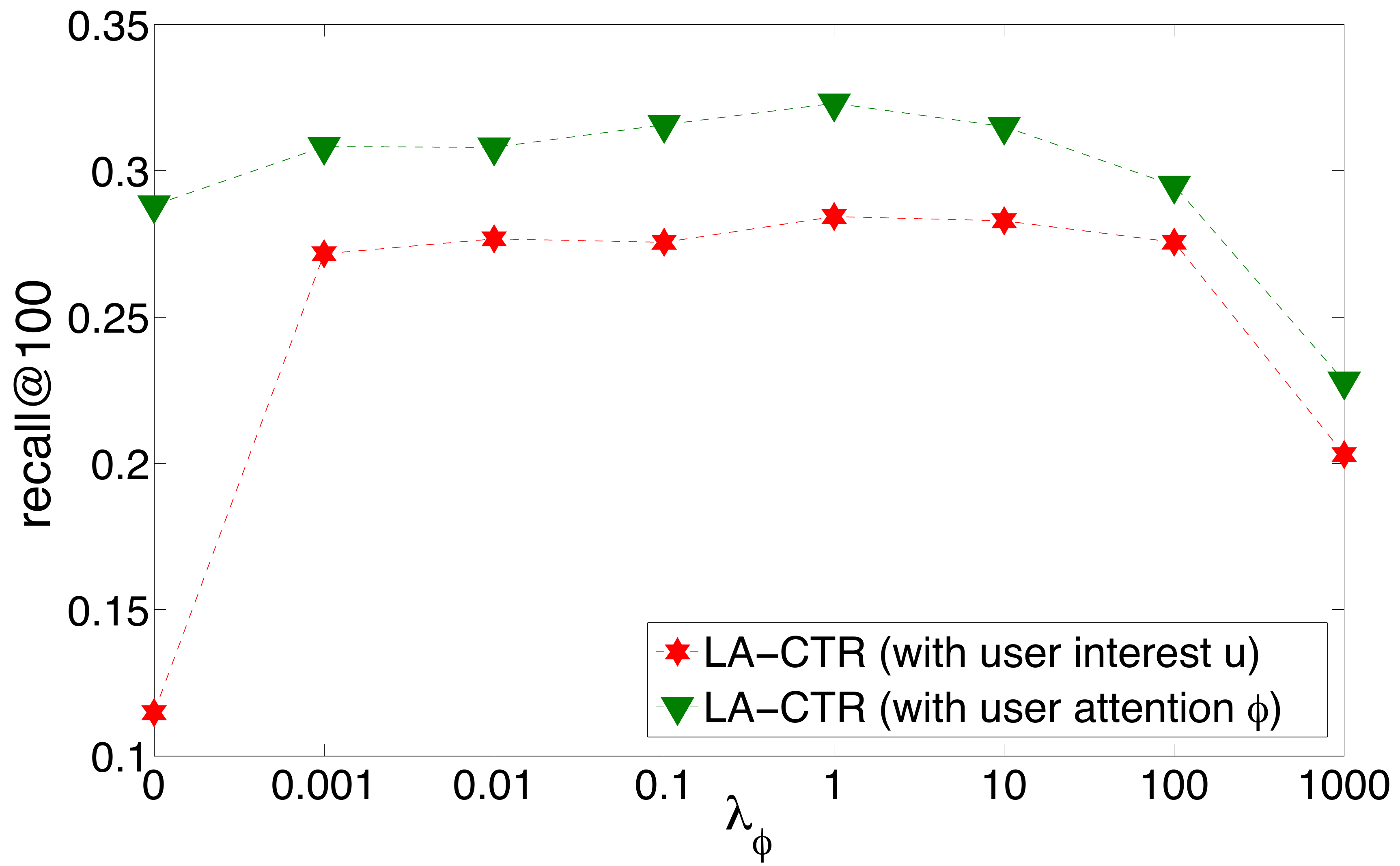}\\
\end{tabular}
\end{center}
\caption{Average recall@\emph{100} of in-matrix prediction for Digg 2009 by varying precision parameters $\lambda_\phi$ while other parameters are fixed.}
\label{fig:parameters}
\end{figure}

We study how parameters of {\laCTR}  affect the overall performance of in-matrix prediction using $recall@100$. The impact of parameter $\lambda_v$ has been studied by CTR and CTR-smf papers, and we found similar trends. Here, we show how the attention parameter ($\lambda_\phi$) affects the overall performance of {\laCTR}. We vary $\lambda_\phi$ $\in$ $\{0, 0.001, ..., 1000\}$, while we fix other parameters: $\lambda_v=100$, $\lambda_u=0.01$, and $\lambda_s=0.01$. A larger value of  $\lambda_\phi$ increases the penalty of user's attention ($\phi$) diverging from interest ($u$) and influence ($s$). In \figref{fig:parameters}, we show the impact of $\lambda_\phi$ on {\laCTR} performance with only user interests ($u$) and with attention ($\phi$).  With small values of $\lambda_\phi$, user's interest ($u$) and influence ($s$) highly diverge from his attention ($\phi$). It explains why $recall@100$ with small $\lambda_\phi$ provides poor performance on recommendation tasks with user interests ($u$) only. Similarly large values of $\lambda_\phi$ will increase the penalty of user's attention diverging from user's influence and interests. Best performance is achieved when $\lambda_\phi=1.0$, for both recommendations with user interest ($u$) and user attention ($\phi$).

\begin{figure}[tbh]
\begin{center}
\begin{tabular}{c}
\includegraphics[width=0.8\linewidth]{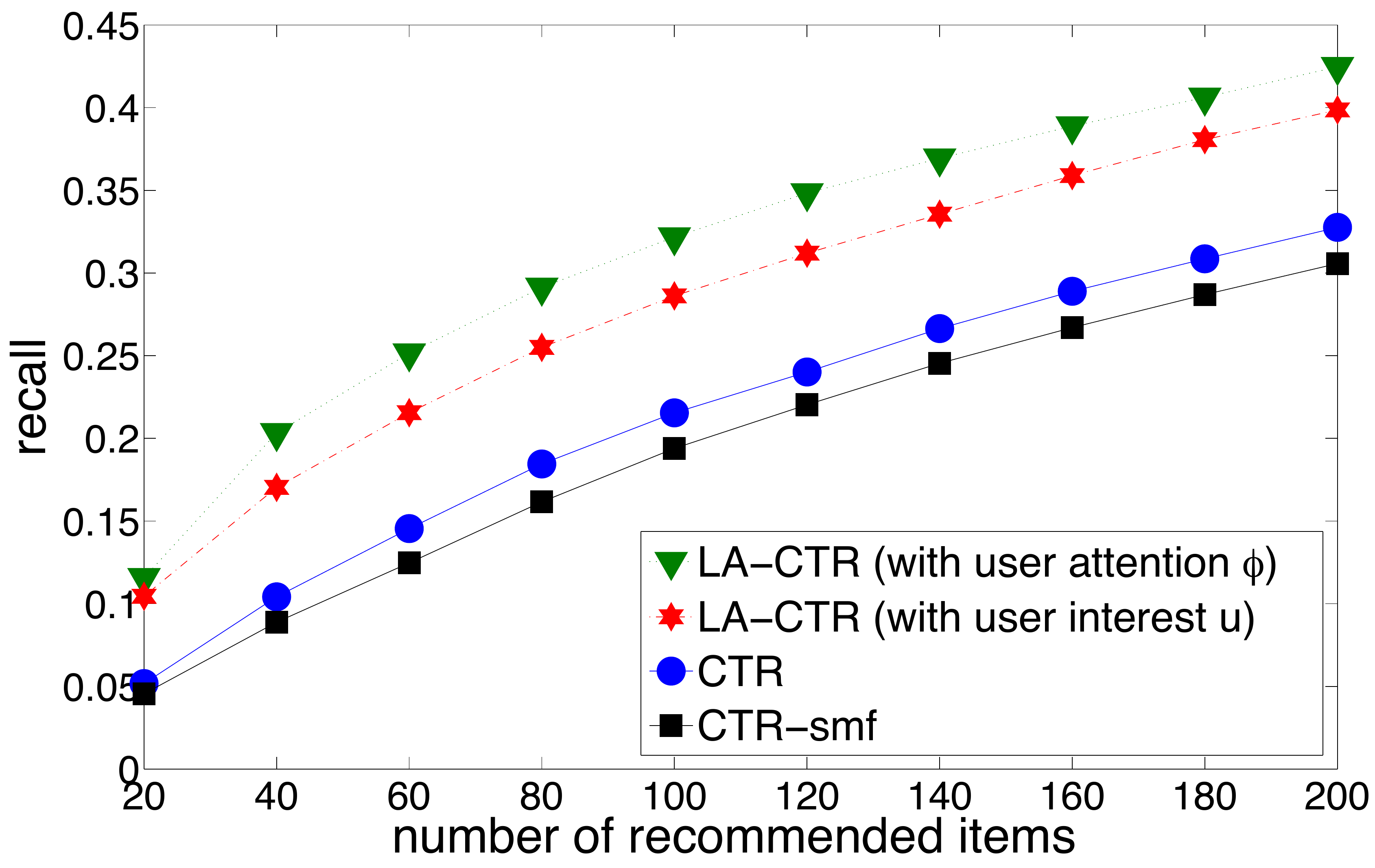}\\
(a)\\
\includegraphics[width=0.8\linewidth]{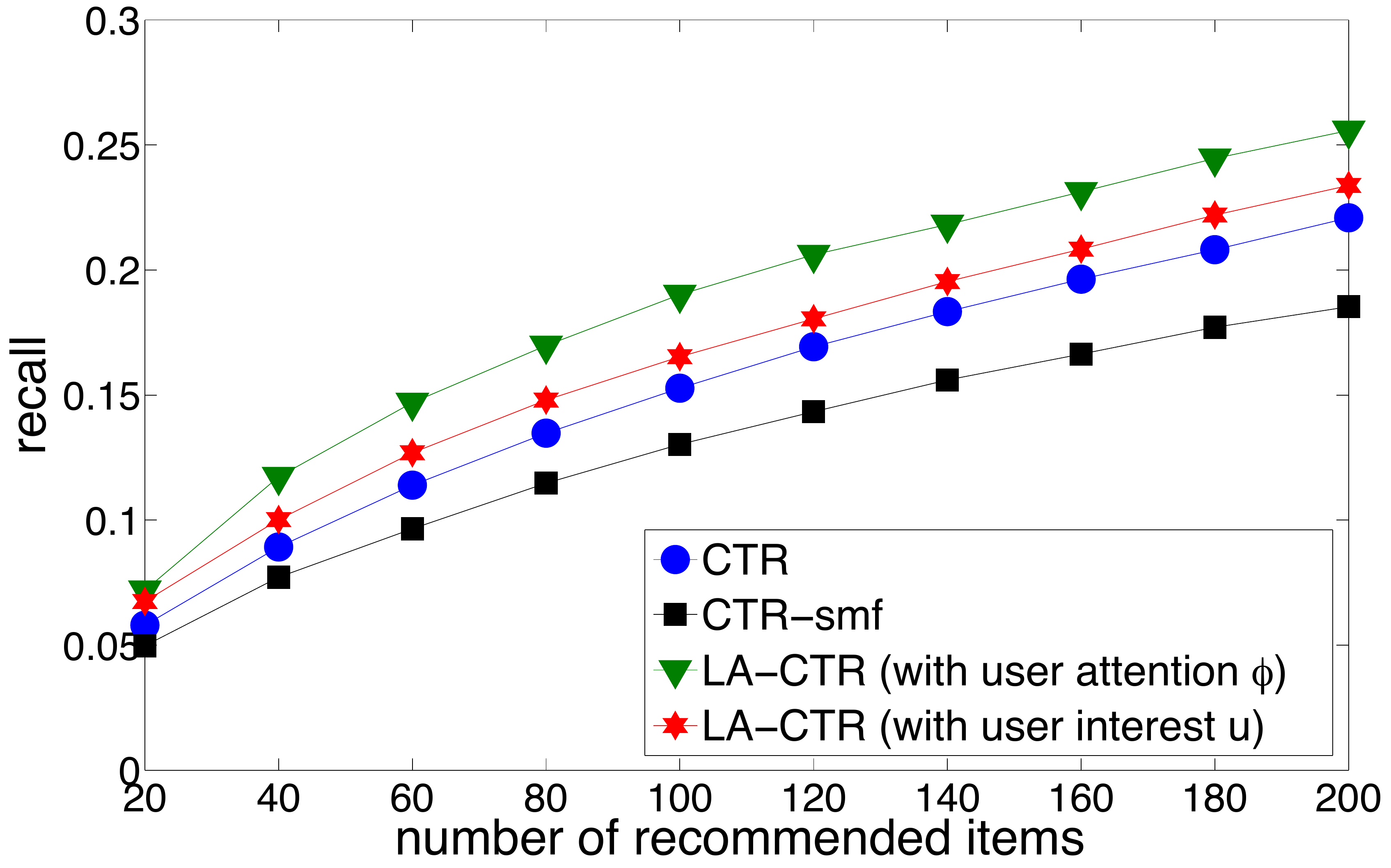}\\
(b)
\end{tabular}
\end{center}
\caption{Recall of in-matrix prediction for (a) Digg 2009 and (b) 2010 data set by varying the number of recommended items (@\emph{X}) with the 200 topics. }
\label{fig:recalldigg0910}
\end{figure}

\subsection{Model Selection}
For collaborative topic regression (CTR), we set the parameters K=200, $\lambda_u $ = 0.01, $\lambda_v $ = 100, a = 1, b = 0.01 by using grid search on held out recommendations. The precision parameters $\lambda_v$ balances how the item's latent vector $v_j$ diverges from the topic proportions $\theta_j$. We vary $\lambda_v$  $\in$ \{0.001, 0.01, 0.1, 1, 10,100,1000\}, where a larger $\lambda_v$  increases the penalty of $v_j$ diverging from $\theta_j$. For Collaborative Topic Regression with social matrix factorization (CTR-smf), we choose the parameters using grid search on held out recommendations. Through out this paper, we set parameters K=200, $\lambda_u $ = 0.01, $\lambda_v $ = 100, $\lambda_q $ = 10,$\lambda_s $ = 2, a=1, and b=0.01.

For {\laCTR}, we choose the parameters similar to CTR and CTR-smf approach by using grid search on held out recommendations. We set parameters K=200, $\lambda_u $ = 0.01, $\lambda_v $ = 100, $\lambda_\phi $ = 1,$\lambda_s $ = 0.01, $a^\phi$=1,$a^r$=1, $b^\phi$=0.01, and $b^r$=0.01.

\subsection{Vote Prediction}
Next, we evaluate the proposed {\laCTR} model by measuring how well it allows us to predict individual votes. There are 257K pre-promotion votes in the 2009 dataset and 1.5 million votes in the 2010 dataset, with 72.34 and 68.20 average votes per story, respectively.  For our evaluation, we randomly split the data into training and test sets, and performed five-fold cross validation. The average percentage of positive examples in the test set is 0.73\%, suggesting that friends share many stories that users end up not voting for. High sparsity of dataset makes the prediction task extremely challenging, with less than one in a hundred chance of successfully predicting votes if stories are picked randomly.

We train the models on the data with selected models of CTR, CTR-smf, {\laCTR} and perform in-matrix prediction  while we vary the number of top $\emph{X}$ items \{20, 40, ..., 200\}. \figref{fig:recalldigg0910} shows the $recall@\emph{X}$ for in-matrix prediction for CTR, CTR-smf, {\laCTR} both with interest and attention. All four models show performance improvement when the number of returned items is increased. Our model with attention always outperformed other three models consistently. Even with learned user interests, {\laCTR} model always outperformed CTR and CTR-smf. Note that recall scores with smaller than 100 of {\laCTR} with user attention are already higher than ones of CTR-smf $recall@200$.
\begin{figure}[tbh]
\begin{center}
\begin{tabular}{c}
\includegraphics[width=0.9\linewidth]{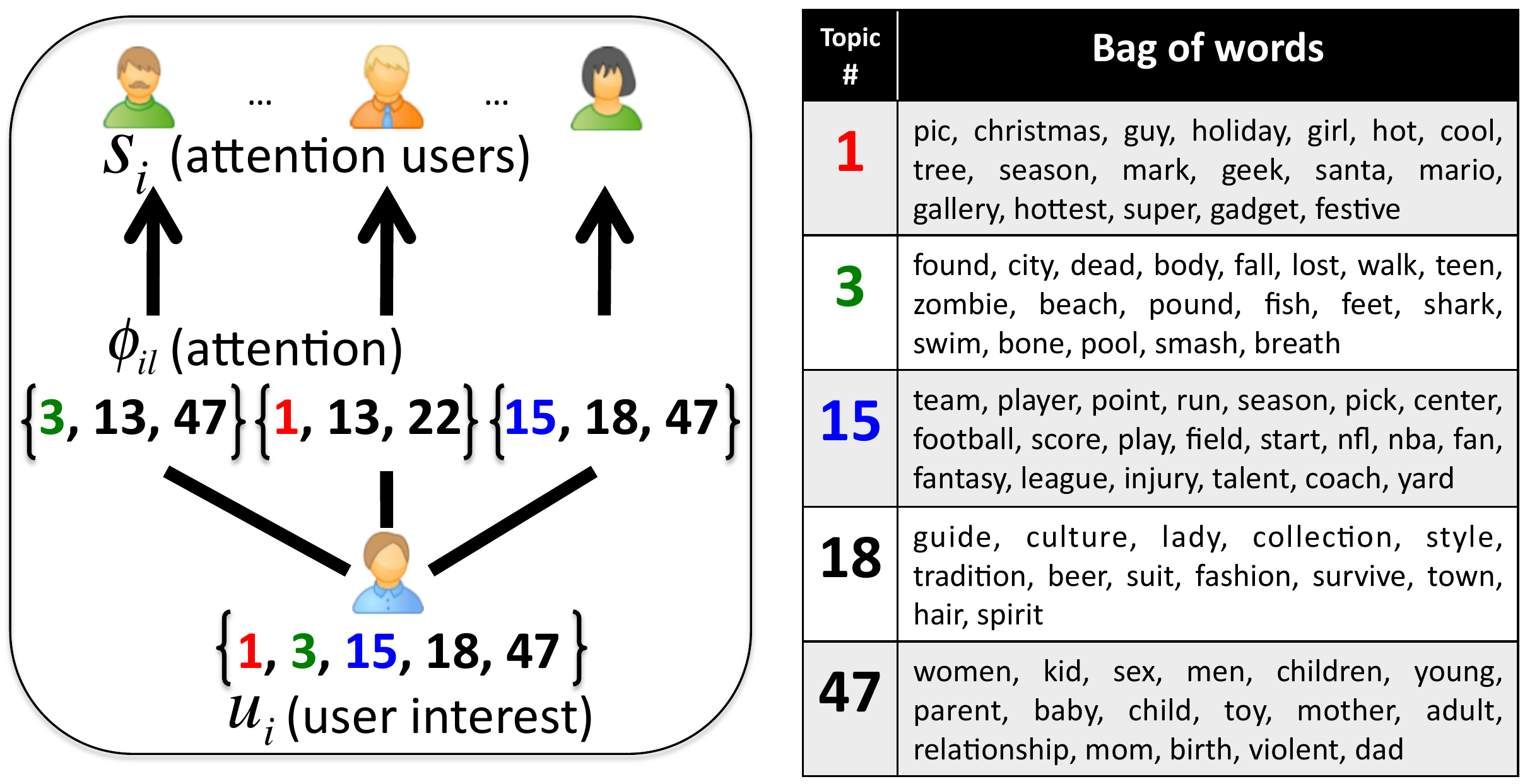}\\
\end{tabular}
\end{center}
\caption{One example user from Digg 2009 data set. We show top 5 topics in learned $u_i$ and top 3 topics in learned attention $\phi_{il}$ to his three most influential users. We also show the bag of words that represents each topic. }
\label{fig:LA}
\end{figure}

\subsection{Limited Attention}
An important advantage of {\laCTR} is that it can explain both course-grained and fine-grained levels of latent topic spaces: (1) user interest topic space ($u_i$) and (2) attention topic space ($\phi_{il}$) using the topics learned from the corpus. Furthermore, it also explains the influence of others ($s_i$) on user's item adoption decisions. In \figref{fig:LA}, we show one example from the 2009 data set. With {\laCTR} we can find the top $n$ matched topics by ranking the elements of his interest vector $u_i$. In this example, user $i$ is interested in topics 1, 3, 15, 18, and 47. Furthermore, we can also explain influence weight of others on user $i$ as well as user $i$'s attention on user $l$ over topic space. Here, user $i$ pays attention to the last user mostly on sports topic (topic \# 15), while he pays attention the first two users on other topics.
\begin{figure}[tbh]
\begin{center}
\begin{tabular}{c}
\includegraphics[width=0.8\linewidth]{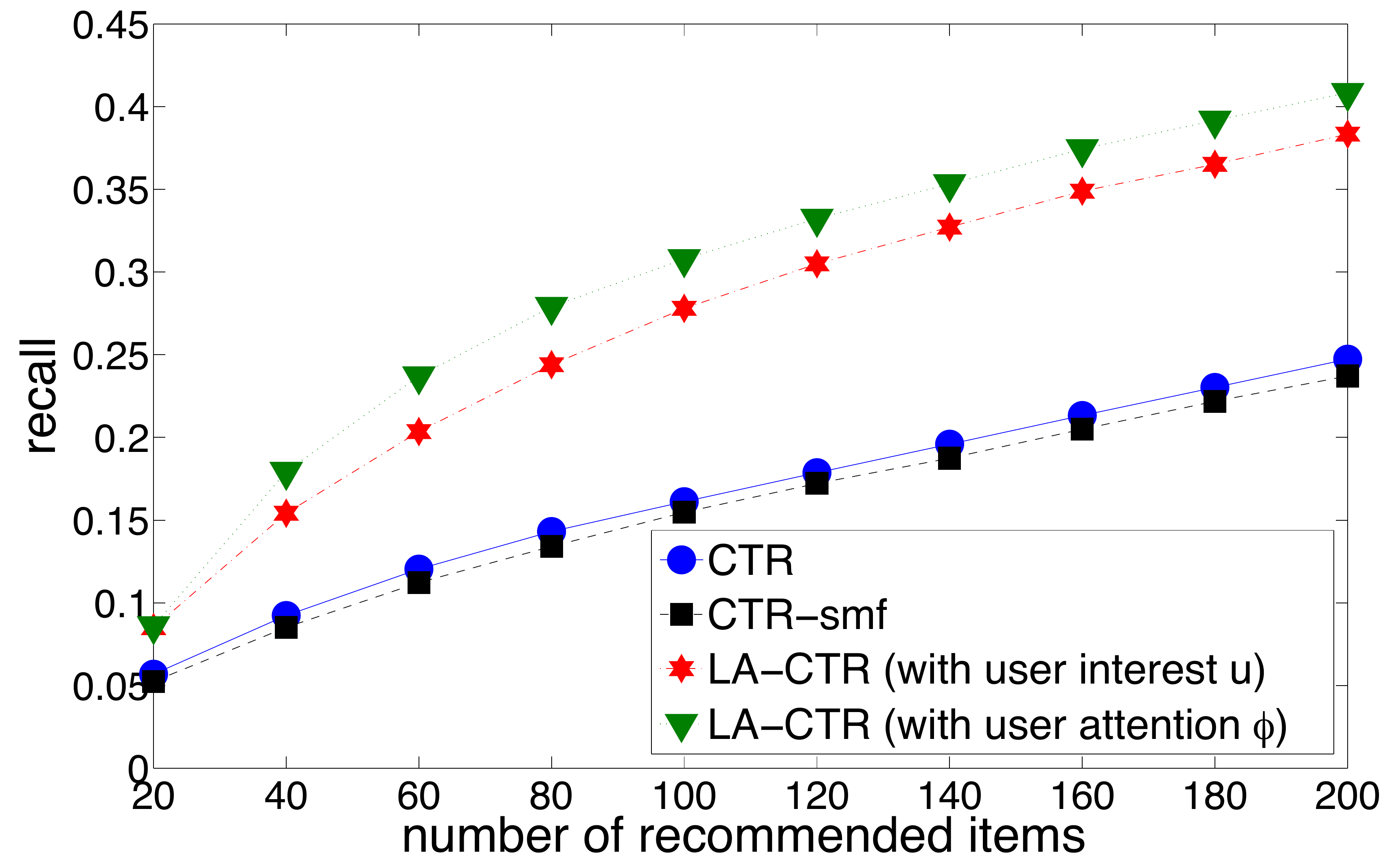}\\
\end{tabular}
\end{center}
\caption{Recall of in-matrix prediction for Digg 2009 by varying the number of recommended items (@\emph{X}) with the 50 topics.}
\label{fig:recalldigg2009with50}
\end{figure}

\section{Conclusion}
In this paper, we proposed {\laCTR} for recommending items  to users in online social media based on users' limited attention as well as their ratings, their social network, and the content of items. We showed that by taking into account users' attention, the proposed model outperforms other state-of-art algorithms on item adoption prediction task. Our model provides not only interpretable user profiles, but also fine-grained description of how much attention the user pays to others on which topics. Such description could help explain why people follow others and how information propagates over the online social network.

One disadvantage of {\laCTR} is its model complexity, since the size of user-item rating is $N^2D$. However, due to high sparsity of the social network and the user-item matrix we did not experience any major slow down. Furthermore, given that with the small number of topics, {\laCTR} already outperformed the other models. In \figref{fig:recalldigg2009with50}, we show recall of in-matrix prediction for Digg 2009 data set with 50 topics. Not only does {\laCTR} model outperform both CTR and CTR-smf with 50 topics, but it also outperformed both CTR and CTR-smf with 200 topics. In other words, increasing the size of the topic space and descriptive power of competitor models did not lead to better performance compared to {\laCTR}. In the future, we will optimize {\laCTR} speed by applying fast bayesian posterior sampling (i.e. stochastic gradient fisher scoring) and will apply it to larger data sets.

For future work, we are interested in applying our model to online social media data set with actual source of adoption information. We will test whether we are able to learn usersÕ limited attention and their interest latent vector better when the actual source of adoption information is available. Note that, our model not only predicts the most likely items but also predicts the most likely source of adoption. With the available source of adoption data set, we will also test our source of adoption predictions for the new item adoptions.

\section*{Acknowledgment}
This material is based upon work supported by  the Air Force Office of Scientific Research under Contract Nos. FA9550-10-1-0102 and FA9550-10-1-0569, and by DARPA under Contract No. W911NF-12-1-0034.

\newpage
\small
\balance
\bibliography{reference}
\end{document}